\documentclass[aps,prl,groupedaddress]{revtex4} \usepackage{graphicx}

\begin{document}

\newcommand{\ba}{\begin{eqnarray}}
\newcommand{\ea}{\end{eqnarray}}
\newcommand{\nn}{\nonumber}
\renewcommand{\d}{\textrm{d}}
\newcommand{\D}{\textrm{D}}
\newcommand{\e}{\textrm{e}}
\renewcommand{\i}{\textrm{i}}
\newcommand{\p}{\partial}
\newcommand{\eps}{\epsilon}
\newcommand{\f}{\phi}
\newcommand{\ve}{\varepsilon}
\newcommand{\cL}{{\cal L}}
\newcommand{\tr}{\textrm{tr}}
\newcommand{\la}{\langle}
\newcommand{\ra}{\rangle}

\newcommand{\fr}[2]{{\textstyle{\frac{#1}{#2}}}}

\title{On the existence of finite-energy lumps in classic field
theories}

\author{Roman V. Buniy}
\email{roman.buniy@vanderbilt.edu}
\affiliation{Vanderbilt University, Nashville, TN 37235}
\author{Thomas W. Kephart}
\email{kephartt@ctrvax.vanderbilt.edu}
\affiliation{Vanderbilt University, Nashville, TN 37235}
\date{\today}

\begin{abstract}
We show how the existence of non-trivial finite-energy time-dependent
classical lumps is restricted by a generalized virial theorem. For
simple model Lagrangians, bounds on energies follow.
\end{abstract}


\maketitle

\section{Introduction} 

Obtaining exact solutions to most realistic field theories is a
formidable task. However, limited information about such solutions is
often available without solving the corresponding equations. One such
general result is Derrick's theorem~\cite{Derrick:1964ww}, which
precludes non-trivial static scalar field configurations in more than
two space dimensions. Some theories have another general feature of
their solutions being classified by their topological charges.

Several non-existence theorems have been proved too. For example, pure
Yang-Mills fields do not hold themselves together to give
finite-energy solutions that are either
time-independent~\cite{Coleman,Deser:1976wq} or periodic in time
\cite{Pagels:1977ck}. An even stronger result was proved: the only
finite-energy non-singular non-radiating solutions with arbitrary time
dependence are vacuum solutions~\cite{Coleman:1977hd}. This ``no-go''
theorem forbids the existence of classical glueballs in a pure
Yang-Mills system. To support localized solutions, other fields have
to be added.

Our particular interest~\cite{Buniy:2002yx} is in glueballs in QCD,
where the $SU_C(3)$ gauge fields couple to quarks and confinement is
involved.  This work is a zeroth order analysis towards understanding
this complete physical situation.

In this note we investigate conditions which are imposed on the
energy-momentum tensor by the existence of such classical solutions
(solitons or lumps). While these solitons are expected to be highly
complicated objects, we do not address their existence nor attempt to
find their explicit forms, but merely find the resulting restrictions
on fields. This information can be used, for example, to find the
lower bound for the energy of lumps as we demonstrate for two
physically relevant systems: scalar fields and scalar fields coupled
to gauge fields. Our theorem is a general result independent of a
particular model, and subject to only mild requirements imposed on
the fields.

\section{Theorem} 

Let us consider a classical field theory in $(n+1)$-dimensional
space-time~\footnote{In QCD or the Standard Model $\cL_m$ depends on
vectors, scalars, and spin-$\frac{1}{2}$ fermions in such a way that
$\cL$ is renormalizable. However, the results given here are
classical, so we need not require renormalizability.} characterized by
the energy-momentum tensor $\theta^{\mu\nu}$ and define a quantity \ba
G^{i\mu}(t,R)=\int_{r\le R}\d^nx\,x^i\theta^{0\mu},\label{G}\ea where
$\mu=0,\ldots n$; $i=1,\ldots,n$ and $r=|{\mathbf x}|$. Using
conservation of the energy-momentum $\p_\mu\theta^{\mu\nu}=0$, we find
\ba\p_0 G^{i\mu}(t,R)=\int_{r\le
R}\d^nx\,\theta^{i\mu}-\int_{r=R}\d^{n-1}
S_jx^i\theta^{j\mu}.\label{dG}\ea We specify the following asymptotic
condition for the energy-momentum tensor,
\ba\lim_{r\to\infty}r^{n+\delta}\,\theta^{\mu\nu}(t,{\mathbf
x})=0.\label{limit}\ea For the energy of the system to be finite we
need $\delta\ge 0$; this also ensures that the surface term in
Eq.~(\ref{dG}) vanishes for large $R$. However, for the quantity
$G^{i\mu}(t,\infty)$ to be finite we need $\delta\ge 1$. If the
function $G^{i\mu}(t,R)$ is \emph{bounded}, its derivative's average
value over a time interval $T$, \ba\la \p_0
G^{i\mu}\ra_T=\frac{1}{T}\int_0^T\d t\,\p_0
G^{i\mu}(t,R)=\frac{1}{T}[G^{i\mu}(T,R)-G^{i\mu}(0,R)],\ea tends to
zero as $T\to\infty$~\cite{Landau}. Averaging Eq.~(\ref{G}) over
infinite time interval and taking the limit $R\to\infty$, we find \ba
\int\d^nx\,\la\theta^{i\mu}\ra=0.\label{K0}\ea

Unfortunately, the above argument fails if $G^{i\mu}(t,R)$ is
\emph{unbounded}. This is a case for many interesting physical
situations, where the energy-momentum goes like $r^{-n-1}$ for large
$r$. For example, the above argument cannot be used for time-dependent
fields that asymptotically go like fields of monopoles or dyons in 4D,
since under the assumption (\ref{limit}), $G^{i\mu}(t,R)$ diverges as
$R^{1-\delta}$ for large $R$. In what follows we will show how the
result (\ref{K0}) can still be proved in such situations.

For now we restrict our attention to lumps at rest; we can choose a
frame such that this is the case for any lump moving slower than the
speed of light. For localized and non-radiating lumps at rest, the
energy-momentum must have an asymptotic behavior (\ref{limit}) with
$0<\delta<1$, which is \emph{uniform in time}~\footnote{For a lump at
rest, outside of a sphere with the radius \emph{independent} of time,
fields approach their asymptotic values with a given accuracy for
\emph{all} times. This guarantees absence of the outgoing
radiation~\cite{Coleman:1977hd}. For a lucid discussion of uniform
convergence see R.~Courant, \emph{Differential and Integral Calculus}
(Interscience, New York, 1937).}.

As a result of these assumptions, for a sphere of large radius $R$,
the surface term vanishes and the right-hand side of Eq.~(\ref{dG})
goes uniformly in time to \ba
K^{i\mu}(t)=\int\d^nx\,\theta^{i\mu}.\label{K}\ea This means that for
any positive $\epsilon$ there exists a \emph{time-independent} $R_0$
such that for any $R>R_0$ we have $\left|\p_0
G^{i\mu}(t,R)-K^{i\mu}(t)\right|<\epsilon$ for all $t\ge 0$. For such
$R$, \ba\left|G^{i\mu}(t,R)-G^{i\mu}(0,R)-t\langle
K^{i\mu}\rangle_t\right|<\epsilon t.\label{ineq1}\ea

On the other hand, using the relation $|{\mathbf P}|\le E$ between the
momentum and energy of the system, we deduce that $G^{i\mu}(t,R)$ is
bounded, \ba|G^{i\mu}(t,R)|\le R\max_{t\ge
0}E(t,R)=\Delta(R);\label{ineq2}\ea $E(t,R)$ being the energy inside
the sphere of radius $R$.

\begin{figure}[tb]
\includegraphics{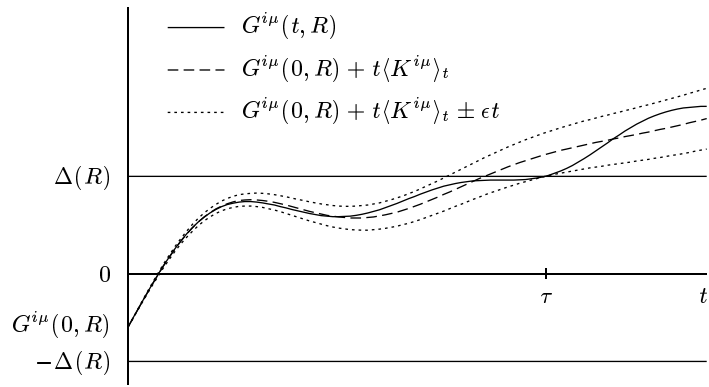}
\caption{For $t>\tau$ the bound (\ref{ineq1}) is valid and
(\ref{ineq2}) is not.\label{figure}}
\end{figure}

Without a detailed form for the function $K^{i\mu}(t)$ we cannot
conclude whether the bounds imposed on the function $G^{i\mu}(t,R)$
are consistent or inconsistent. There is, however, a simple case where
inconsistency is obvious. Let the average $\langle K^{i\mu}\rangle$ be
not zero. Since $\epsilon$ is arbitrary, we can choose
$\epsilon<|\langle K^{i\mu}\rangle|$. Then bounds (\ref{ineq1}) and
(\ref{ineq2}) cannot possibly be consistent for all $t\ge 0$ (see
Figure~\ref{figure}). It follows that the only resolution is to set
$\langle K^{i\mu}\rangle=0$; this gives Eq.~(\ref{K0}).

For pure Yang-Mills theory in 4D, $\langle {K^i}_i\rangle=0$ leads to
$E=0$, which is Coleman's conclusion in
Ref.~\cite{Coleman:1977hd}. One can imagine several cases where
$\langle K^{i\mu}\rangle=0$ is fulfilled: (1) $K^{i\mu}(t)$ tends to
zero as $t$ goes to infinity; (2) for large $t$, the function
$K^{i\mu}(t)$ approaches a periodic function that oscillates around
zero.

We conclude this section with a remark on a system described by the
Lagrangian density $\cL(\phi,\p_\mu\phi)$. The dilatation
transformation $\delta\phi=x^\mu\p_\mu\phi$ of such density is
\ba\delta\cL=\p_\mu(x^\mu\cL)+{\theta^\mu}_\mu.\ea From
Eq.~(\ref{K0}), it follows that the time average of the Lagrangian is
the system's energy, $\la\delta L\ra=E$.

\section{Examples}

We apply our result to two theories for which lumps exist (for a
review see e.g. \cite{Rajaraman}).

\emph{1.---} For a scalar field theory with the Lagrangian density
\ba\cL=\fr{1}{2}\p_\mu\phi_a\p^\mu\phi_a-U(\phi_a),\ea the requirement
$\int\d^nx\,\la{\theta^i}_i\ra=0$ gives the following expression for
the energy, \ba E=\int\d^nx\,\la
2U(\phi_a)+(1-\fr{1}{n})(\p_i\phi_a)^2\ra.\label{EU1}\ea We thus have
a bound \ba E\ge 2\int\d^nx\,\la U(\phi_a)\ra.\label{bound}\ea For one
space dimension the energy density becomes $\la 2U(\phi)\ra$. This
agrees with a well-known result for a static scalar field with one
component.

For a single scalar field in one space dimension, the potential
$U(\phi)=1-\cos\phi$ leads to the sine-Gordon equation, which has a
breather solution \ba\phi(t,x)=4\tan^{-1}\left[
\frac{\sqrt{1-\omega^2}\sin{\omega t}}
{\omega\cosh{x\sqrt{1-\omega^2}}}\right].\ea It can be readily checked
that this solution saturates the bound (\ref{bound}).

\emph{2.---} For a theory of coupled scalar and gauge fields with the
Lagrangian density \ba\cL=-\fr{1}{4}F^{\mu\nu}_a
F^a_{\mu\nu}+\fr{1}{2}\D_\mu\phi_a\D^\mu\phi_a-U(\phi_a)\label{Aphi}\ea
we similarly find the total energy is \ba E=\int\d^nx\,\la
2U(\phi_a)+(1-\fr{1}{n})(\D_i\phi_a)^2+(\fr{2}{n}-1)F_{0i}^aF_a^{0i}
+\fr{1}{2n}F_{ij}^a F_a^{ij}\ra.
\label{EU2}\ea For two and three spacial dimensions (the only
interesting cases for the Lagrangian (\ref{Aphi})) the sum of last
three terms on the right-hand-side in Eq.~(\ref{EU2}) is positive. We
thus have a bound \ba E\ge2\int\d^nx\,\la U(\phi_a)\ra,\ \ \ \ \
n=2,3.\ea

\section{Massless lumps}

We now turn to the case of lumps moving with the speed of light. We
carry calculations for a specific model, the 4D Yang-Mills sytem with
sources, modifying the argument in Ref.~\cite{Coleman:1977hd}.

By choosing the 3-axis in the direction of the momentum, we make the
fields transverse with their components related by
$E_\alpha=\epsilon_{\alpha\beta}H_\beta$. In terms of light-cone
variables $x^{\pm}=x^0\pm x^3$, the only non-vanishing components of
the field-strength are $F_{+\alpha}=-2E_\alpha$. Since $F_{12}=0$, we
can perform a gauge transformation depending on $x^1$ and $x^2$ to set
$A_1$ and $A_2$ to zero. From $F_{-\alpha}=0$ it now follows that
$A_-$ is independent of $x^1$ and $x^2$, so we make a gauge
transformation depending on $x^-$ to set $A_-=0$. Next, from the
Yang-Mills equations of motion with sources $J_\mu$ we find $J_{-}=0$
and \ba&&\p^\alpha\p_\alpha A_+=J_{+},\label{eq1}\\&&\p^+\p_\alpha
A_++[A^+,\p_\alpha A_+]=-J_{\alpha}.\label{eq2}\ea Eq.~(\ref{eq2})
follows from Eq.~(\ref{eq1}) by differentiation and using covariant
conservation of the current. We are left with only Eq.~(\ref{eq1}) to
solve and its general solution is \ba
A_+(x^+,x^1,x^2)=\tilde{A}_+(x^+,x^1,x^2)
+\frac{1}{4\pi}\int\d\xi^1\d\xi^2\,J_{+}(x^+,\xi^1,\xi^2)
\log\left[{(x^1-\xi^1)}^2+{(x^2-\xi^2)}^2\right],
\label{Poisson}\ea where $\tilde{A}_+$ is a solution 
to the Laplace equation $(\p^1\p_1+\p^2\p_2)\tilde{A}_+=0$. It is well
known that the only non-singular solution to this equation is a
function $\tilde{A}_+(x^+)$ of $x^+$ alone. For a non-singular
current, the second term in Eq.~(\ref{Poisson}) can have a singularity
only at infinity. For large $r$, the second term in
Eq.~(\ref{Poisson}) is asymptotically $\frac{1}{2\pi}I\log{r}$, where
\ba
I=\int\d\xi^1\d\xi^2\,J_{+}=\oint_C(F_{+2}\d\xi^1-F_{+1}\d\xi^2),\ea
and $C$ is an infinite contour in the $(\xi^1,\xi^2)$ plane, which
encloses the sources. Observe that the contour integral form for $I$
is the gauge field flux in the transverse plane. Fields are
non-singular only when the transverse flux vanishes, $I=0$; the
asymptotic $r^{-2}$ behavior of the field-strength ensures this. In
contrast to the relativistic source-free case where only vacuum
solutions exist, here we can only constrain the form of lumps moving
with the speed of light.

\section{Conclusions}

To summarize, we have proved a generalized virial theorem which
restricts possible forms of massive classical lumps. In particular, it
establishes the lower energy bounds for such objects. Also, we have
found only mild restrictions on massless solutions. These conclusions
were reached for classical systems and they do not restrict forms of
possible quantum lumps.

\bibliography{lumps}

\end{document}